\documentclass[a4paper,fleqn,usenatbib]{mnras}

\usepackage{newtxtext,newtxmath}

\usepackage[T1]{fontenc}
\usepackage{ae,aecompl}

\usepackage{graphicx}	
\usepackage{amsmath}	
\usepackage{amssymb}	
\usepackage{gensymb}

\title[Ultraviolet Extinction in Orion]{Interstellar extinction in Orion.\\ 
Variation of the strength of the UV bump across the complex.  }

\author[L. Beitia-Antero  and A. I. G\'omez de Castro]{
Leire Beitia-Antero,$^{1}$\thanks{E-mail: lbeitia@ucm.es}
Ana I. G\'omez de Castro,$^{1}$\thanks{E-mail: aig@ucm.es}
\\
$^{1}$AEGORA Research Group, Universidad Complutense de Madrid,
Facultad de CC. Matem\'aticas, Plaza de Ciencias 3, E-28040 Madrid
}

\date{Accepted XXX. Received YYY; in original form ZZZ}

\pubyear{2017}

\begin{document}
\label{firstpage}
\pagerange{\pageref{firstpage}--\pageref{lastpage}}
\maketitle

\begin{abstract}

There is growing observational evidence of dust coagulation in the dense filaments within molecular clouds. 
Infrared observations show that the dust grains size distribution gets shallower and the relative 
fraction of small to large dust grains decreases as the local density increases. Ultraviolet (UV)
observations show that the strength of the $2175~\text{\AA}$ feature, the so-called UV bump, also 
decreases with cloud density.  In this work, we apply the technique 
developed for the Taurus study to the Orion molecular cloud and confirm that the UV bump 
decreases over the densest cores of the cloud as well as 
in the heavily UV irradiated  $\lambda$ Orionis shell. The study has been extended to the Rosette cloud
with uncertain results given the distance (1.3 kpc).

\end{abstract}

\begin{keywords}
ISM: dust, extinction
\end{keywords}



\section{Introduction}\label{sec:introd}
The Interstellar Medium (ISM) is a complex ensemble of gas, large molecules and dust coupled 
through the action of the magnetic fields permeating the Galaxy.  In some parts of the Galaxy,
molecular clouds are formed due to galactic stresses (\cite{Elmegreen90,Franco02}) or eventual 
accretion from the intergalactic medium (\cite{Mirabel82,Wang04}). In the outskirts of these 
clouds the dust size distribution varies due to density gradients 
that determine the characteristics of the stars
that will be formed inside them (see e.g. \cite{Dzyurkevich2017}). Particularly,
small dust grains are relevant because they act as main carriers of the field and
set the fundamental scales for ambipolar diffusion and magnetic waves
propagation (\cite{Pilipp87}), favouring the coupling of the cloud
with the ambient field up to cut-off wavelengths
of 0.1pc (\cite{Nakano98}).\par

 The distribution of large dust grains has been studied, for instance, by The Herschel
Space Telescope. It 
has provided measurements of this population in the nearest molecular complexes such as 
Taurus (\cite{Ysard2013}) and Orion  (\cite{Roy2013, Stutz2015}). On the other hand, the most
sensitive way
to detect variations in the distribution of small dust grains (<0.05 microns) 
and large molecules (such as the Polycyclic Aromatic Hydrocarbons or PAHs) is to measure
the strength of the 
$2175~\text{\AA}$ feature of the extinction curve, the so-called UV bump. The UV bump is,
by far, the strongest 
spectral feature in the extinction curve but its source remains uncertain (see {\it e.g.}
the review 
by \cite{Draine03}). Though small graphite grains  were proposed initially as the main
source of the bump, 
the baseline today are PAHs (\cite{WeingartnerDraine01}), that share with a graphite sheet
a similar structure in terms of the distribution of the  carbon atoms. PAHs are required to 
reproduce the observed infrared emission (\cite{LegerPuget1984})  and they somewhat represent 
the extension of the dust grain size distribution into the molecular domain. \par

 The relative fraction of the smallest components of dust grains in the ISM can be modified
through coagulation into larger particles in regions of high density or through 
destruction by a strong ultraviolet radiation field. In a previous work,
we developed a method to explore the former process applying the star counts  method in the
near ultraviolet
(NUV) (\cite{GdC15}, hereafter GdC2015),  using the {\it All Sky Survey} (AIS) carried out 
by the Galaxy Evolution Explorer (GALEX)
The method was successfully applied to the
Taurus molecular complex and
 we found evidence
of a decreasing UV bump strength as approaching to the densest
parts of the clouds, in excellent agreement with
Spitzer-based results (\cite{Flagey09}). \par

In this work, we apply the same method to the Orion molecular 
cloud  to detect possible dust coagulation processes and to explore
the impact of the strong UV background in the PAHs survival. 
Orion  spans more than 700deg$^2$ in the sky and it contains three main molecular
structures: the Orion A ($d = 371\pm 10$pc) and Orion B ($d=398\pm 12$pc) 
clouds and the $\lambda$ Orionis ring ($d=445\pm 50$pc) (distances
from \cite{LAL11}).
In Section \ref{sec:method}, the method is shortly described and in Sections
\ref{sec:data_proc} and \ref{sec:results}, it is applied to 
the Orion star forming region. Evidence of UV bump suppression close to the dense
filaments is found, 
as well as deviations from the ISM law around the heavily irradiated area
  of $\lambda$ Orionis.
Finally, in Section \ref{sec:rosette}, the work has been extended to the
nearby (in projection) Rosette cloud.
Ultraviolet spectroscopic observations are available for some Rosette stars that have been used
to test the statistical quality of the results. A brief summary is provided
in Section \ref{sec:conclusions}.
\section{Method description}\label{sec:method}

The method is based on the extensively used extinction model derived
by \cite{FM07} (hereafter FM07). GdC2015 show that the strength of the bump
can be expressed in terms of extinction in the GALEX NUV band, $A_{\rm NUV}$, 
and in the K  infrared band, $A_{\rm K}$, as, 
\begin{equation}\label{eq:linear_relation}
 A_{\rm bump} = (0.106\pm0.008)\frac{A_{\rm NUV}}{A_{\rm K}} + (2.0\pm 0.3) 
\end{equation}

\noindent
where $A_{\rm bump}$ is the strength of the bump in the extinction curve (FM07).
This result simply indicates that the ISM extinction law admits a fairly simple parametric 
approximation from the near infrared to near ultraviolet except for the strength of the bump. 

$A_K$ has been evaluated for the nearby molecular clouds by \cite{Froebrich07} (hereafter F07)
based on data from the 2MASS survey (\cite{Skrutskie06}).  $A_{\rm NUV}$
may be calculated by using
the star counts method (GdC2015) that has been extensively used in astronomy (\cite{BokCordwell1973}).
For example, this method was applied by \cite{LAL11} to estimate the distances to the main components of the 
Orion complex ($d(Ori~A) = 371\pm 10pc$, $d(Ori~B) = 398\pm12pc$, $d(\lambda ~Ori) = 445\pm 50pc$).

In its most usual application, star counts in a given field are
compared with the predictions for an unextincted, nearby field and the extinction is given by,

\begin{equation}\label{eq:extinction}
A_{\rm NUV}=\frac{1}{b_{\rm NUV}}\log(N^{*}_{\rm NUV}/N_{\rm NUV})
\end{equation}
\noindent
where $N_{\rm NUV}$ are the observed counts, $N^{*}_{\rm NUV}$ are the expected counts from a 
non-extinguished field, and $b_{\rm NUV} = d\log N(m_{\rm NUV})/dm_{\rm NUV}$ 
(with $m_{\rm NUV}$, the apparent magnitude at wavelength $\rm NUV$) is a measure of the 
slope of the luminosity function in the area under study. 
As the NUV luminosity function 
has not been determined for the Galaxy yet, we have derived it for the Orion region using GALEX data.

The resolution of the maps created by these means depends on the angular size of the
region used to measure the star counts that becomes the picture element, the pixel,
of the extinction map. A compromise must be sought between resolution and
good statistics. Large pixel sizes are
ideal for deducing statistical properties but, in practice,
averaging the extinction over a large area smooths the results. Small pixel sizes provide a cleaner idea
of the distribution of the obscuring material but the statistics may become very poor and, with them,
the extinction measurements. Based on our previous experience (GdC2015), we have worked with
three pixel sizes:  30arcmin, 15arcmin and 10arcmin (see
Section \ref{sec:results} for a discussion of the
results). 

\section{Data processing}\label{sec:data_proc}

\subsection{Data}

The GALEX mission was intended to survey the ultraviolet sky avoiding the Galactic Plane,
as well as regions with bright sources that could damage the detectors
(\cite{GALEX}).
GALEX images (or tiles) were obtained  in two UV bands: Far Ultraviolet (FUV) band
and Near Ultraviolet (NUV) band, with a circular field of
view of  $1.2\degree$.
The GALEX AIS was completed in 2007 and
 covers $\sim 26,000$deg$^{2}$. 
Point-like sources were extracted from the images to generate the GALEX catalogue
(\cite{GALEX-GR5}) that is used as  a baseline for this work;  only
NUV data will be
used  due to statistical reasons  since
GALEX detected $\sim 10$ times more sources in 
NUV band than in FUV band.\par

\begin{figure*}
  \includegraphics[width=16cm]{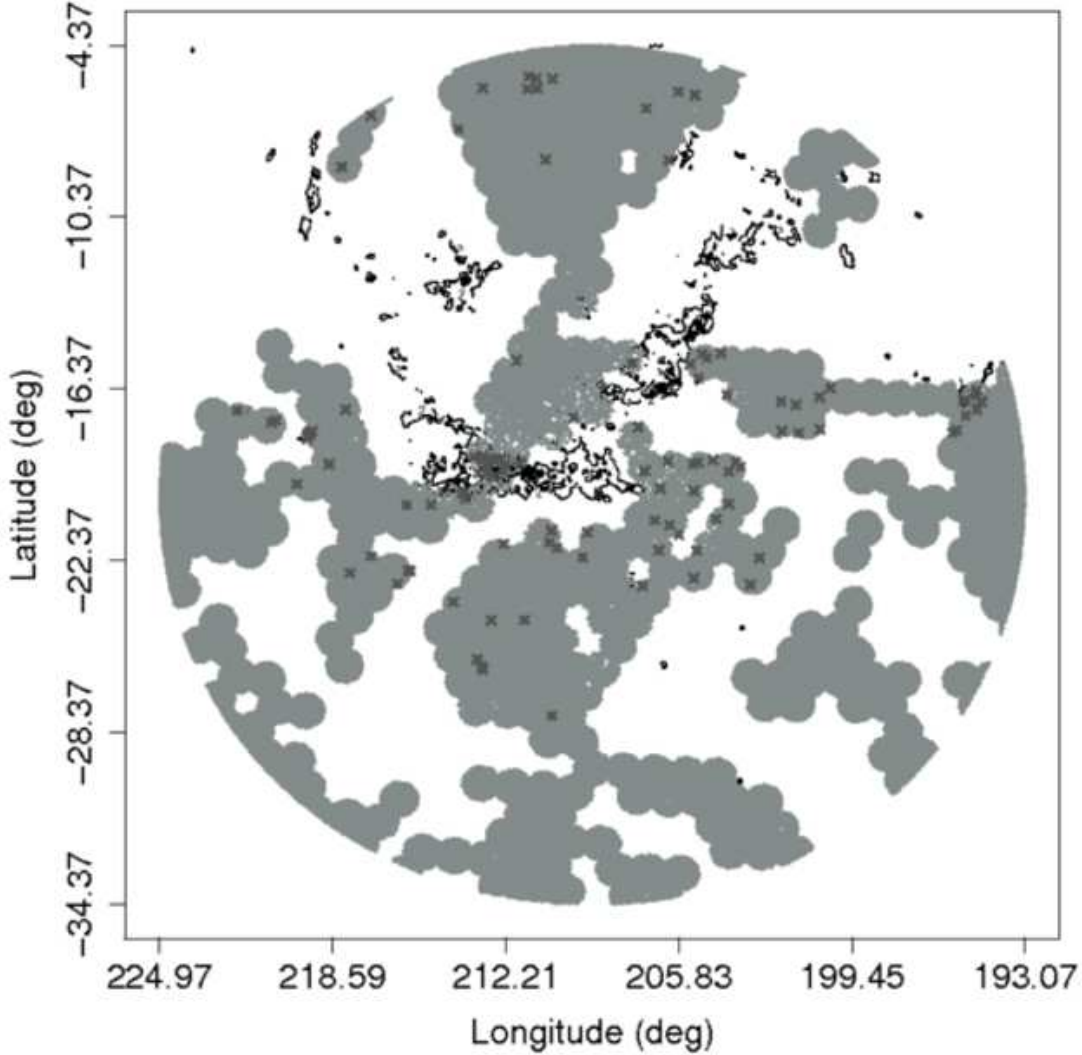}
  \caption{Distribution of the GALEX NUV sources centered in Orion  in a
      search radius of $r=15\degree$ superimposed on the infrared
    dust map by F07. Note that the 0.$\degree$6 radius field of the GALEX tiles is clearly recognisable.  Marked
with a cross are the Young Stellar Object candidates from S2014.}
\label{fig:sources_distr}
\end{figure*}

In the Orion region, the survey consists of 359 images (or tiles): $\sim 406$deg$^{2}$, 
including a small fraction of the Orion A cloud and part of the $\lambda$ Orionis ring. 
To remove spurious sources, the NUV sources in the catalogue were cross-correlated with 
the 2MASS catalogue (see \cite{GdC11, Bianchi11} for details).
Only NUV sources with 2MASS counterpart in a search radius of 3arcmin were 
considered bona-fidae sources for the application of the star counts method (\cite{GdC15}). The
selection of the NUV sources in the Orion region was obtained in the same manner (\cite{Sanchez14}, hereafter S2014);
the final sample amounts to a total of 289,968  NUV sources. Their spatial distribution is
shown in Figure \ref{fig:sources_distr} together with F07's A$_{\rm V}$-map
as a reference. Also overplotted are the 111 Young Stellar Objects candidates determined by S2014;
concentrations are observed towards the Orion A cloud and the $\lambda$ Orionis ring.\par

The sensitivity of the star counts method depends on the vast majority of the stars being background 
to the cloud under study. As the Orion cloud is at $\sim 400$ pc from the Sun, most sources are
expected to be background. The release of the Gaia DR1 catalogue
(\cite{GaiaDR1}) has allowed an assessment of the
fraction of foreground stars. 
Unfortunately, only 13,562 stars have been
found in the GAIA - GALEX cross-match; {\it i.e.} about $4.7$\% of the sample.
33\% of these sources (4,444) 
are foreground stars; this factor is an upper limit since GAIA's first release is not complete neither in brightness 
(it is balanced to nearby sources) nor in direction (\cite{GaiaDR1}). 
Note that the contribution of foreground sources can be neglected if they are
uniformly
distributed over the field and the density of foreground sources is much smaller than the
density of  background sources\footnote{Let be $N_0$ the density of foreground sources, then 
Eq.~2 changes to $A_{\rm NUV}=\frac{1}{b_{\rm NUV}}\log \bigg(\frac{N^{*}_{\rm NUV}+N_0}{N_{\rm NUV}+N_0}\bigg)$ and,
$A_{\rm NUV}=\frac{1}{b_{\rm NUV}}\log \bigg(\frac{1+N^{*}_{\rm NUV}/N_0}{1+N_{\rm NUV}/N_0}\bigg)$.},
which is the case.

\subsection{Calculation of the A$_{\rm NUV}$ map}

In order to evaluate the map it is necessary to determine the slope of the NUV luminosity function
in Orion and to select the reference field for the calculation
of the extinction.
As the area mapped is  406$\degree$  wide, and extends over about 30$\degree$
in galactic latitude and longitude,
some precautions need to be taken. Firstly, possible deviations of the luminosity 
function across the region must be evaluated. Then, as the density of stars decreases with
galactic latitude, it is required to correct for this effect when defining the $N^{*}_{\rm NUV}$
for each galactic latitude.

\subsubsection{NUV Luminosity function}

The region was divided in square regions of $2\degree \times 2\degree$
over which we computed the luminosity function; among them we chose
ten, apparently non-extinguished comparing to IRIS and Planck dust maps.
In each case, we fitted the slope of the luminosity function
in the 16-20 mag interval (the completeness region) by ordinary
least squares obtaining the values in Table \ref{tab:res_lumfunc}.

\begin{table}
  \caption{Least-squares fitting results of the NUV luminosity function
    for the 10 regions in Orion
    in the 16-20mag interval.
    $^{*}S_{r}$: residual standard error.}
  \label{tab:res_lumfunc}
  \centering
  \begin{tabular}{ccccc}
    \hline
    Region & Stars & $b_{\rm NUV}$ & $e_{b_{\rm NUV}}$ & $S_{r}^{*}$\\
    \hline
    R1  & 678 & 0.264 & 0.019 & 0.061 \\
    R2  & 1982 & 0.280 & 0.011 & 0.036 \\
    R3  & 1062 & 0.279 & 0.007 & 0.024\\
    R4  & 971 & 0.303 & 0.016 & 0.051\\
    R5  & 1003 & 0.276 & 0.023 & 0.073\\
    R6  & 443 & 0.292 & 0.024 & 0.078\\
    R7  & 795 & 0.263 & 0.017 & 0.056\\
    R8  & 702 & 0.253 & 0.020 & 0.066\\
    R9  & 671 & 0.253 & 0.019 & 0.062\\
    R10 & 772 & 0.261 & 0.018 & 0.058\\
    \hline
  \end{tabular}
\end{table}

 Then we tested whether there were significant variations with galactic latitude
or longitude and not finding such a trend, we computed a weighted mean
obtaining $b_{\rm NUV} = 0.276\pm 0.004$, which is quite similar to but smaller
than the value obtained in Taurus ($b_{\rm NUV} = 0.294\pm 0.002$, GdC2015).

\subsubsection{Stellar density as a function of galactic latitude}

Stellar density decreases with galactic latitude introducing a bias
in the extinction value that has been corrected following GdC2015. A simple
inspection of Eq.~\ref{eq:extinction} shows that if $N^{*}_{\rm NUV}$ depends on 
galactic latitude, then

\begin{equation}\label{eq:dAnuv1}
A_{\rm NUV}= \frac{1}{b_{\rm NUV}} \log \bigg( \frac{N^{*,b0}_{\rm NUV}}{N^{*,b}_{\rm NUV}}\frac{N^{*,b}_{\rm NUV}}{N_{\rm NUV}} \bigg)
\end{equation}

\noindent
being the true extinction, $A_{\rm NUV}^{0}$ given by,
\begin{equation}\label{eq:dAnuv2}
A_{\rm NUV}^{0}= \frac{1}{b_{\rm NUV}} \log \bigg(\frac{N^{*,b}_{\rm NUV}}{N_{\rm NUV}}\bigg)
\end{equation}

\noindent
hence,

\begin{equation}\label{eq:dAnuv3}
  A_{\rm NUV}^{0}=A_{\rm NUV}-dA_{\rm NUV}
\end{equation}

\noindent
where $dA_{\rm NUV}$ is the correction factor,

\begin{equation}\label{eq:dAnuv4}
dA_{\rm NUV}=  \frac{1}{b_{\rm NUV}} \log \bigg(\frac{N^{*,b0}_{\rm NUV}}{N^{*,b}_{\rm NUV}}\bigg)
\end{equation}

\noindent
that depends on galactic latitude. To evaluate this correction,
we have selected a set of non-extiguished fields and have evaluated $dA_{\rm NUV}$, using
as $N^{*,b0}_{\rm NUV}$ the maximum value; we have also included
a field at $l_{gal}\sim 208\degree.745$, $b_{gal}\sim +0.83\degree$ to reach the 
Galactic Plane. Then, we have fitted these $dA_{\rm NUV}$ values as
a linear function of $b_{gal}$ using the Theil-Sen non-parametric
 fitting method 
(see Appendix \ref{ap:galactic_correction} for details on the final fit).\par

\subsection{The A$_{\rm NUV}$/A$_K$ map}\label{subsec:bump}

Three A$_{\rm NUV}$ maps were computed for pixel sizes 10, 15 and 30 arcmin
after applying the method and corrections described above. The infrared data
were extracted from F07's map based on 2MASS data.

2MASS performed the most complete near infrared survey of the sky providing 
JHK magnitudes for 471 million of point sources. There exist numerous dust maps 
based on 2MASS data (e.g. \cite{LAL11, JuvelaMontillaud2016}). 
We have selected the dust map of the Galactic Anticentre by F07 {because of its
availability in FITS format as online material
that was obtained using the NICE (\textit{Near Infrared Color
Excess}) method developed by \cite{Lada94}. 
As F07 point out, ``extinction values in the very dense regions of the clouds 
in this map are underestimated and there is a small $A_{\rm V}$ dilution
for more distant clouds''. However, as Orion is at a moderate
distance ($\sim 400$pc) and there are not intervening molecular clouds
in the line of sight, any geometric effect will affect
the same way the extincion measures in NUV and K bands, provided that
the pixel size is the same. We could only find underestimated values 
in the small portion of the filament of Orion A that has been observed by GALEX,
which is the densest region in our map.

F07's map was downloaded in FITS format via ftp. We transformed back the F07's $A_{\rm V}$ map
into an $A_{\rm H}$ map following F07 instructions, and then in an A$_K$ map by applying 
the interstellar extinction law by FM07: $A_{\rm H}=1.666A_{\rm K}$. The map was then resampled to the selected resolutions (10', 15' and 30') and the
ratio A$_{\rm NUV}$/A$_{\rm K}$ was computed. The
final images are plotted in Fig. \ref{fig:Orion_panel}\par

In general, at lower latitudes the $A_{\rm K}$ map contains a large
number of negative or zero extinction values due to
the scarce sensitivity of near infrared  wavelengths to small
dust columns, which results in
abnormally high $A_{\rm NUV}/A_{\rm K}$ values. In our maps, the
saturation value has been set at 33, which corresponds to the mean value for the ISM (GdC2015).
In order to discern the relationship between the high column density areas of gas and
the regions of grain growth, we have plotted the relative extinction maps
over the dust map by F07.\par

\begin{figure}
  \includegraphics[width=\columnwidth]{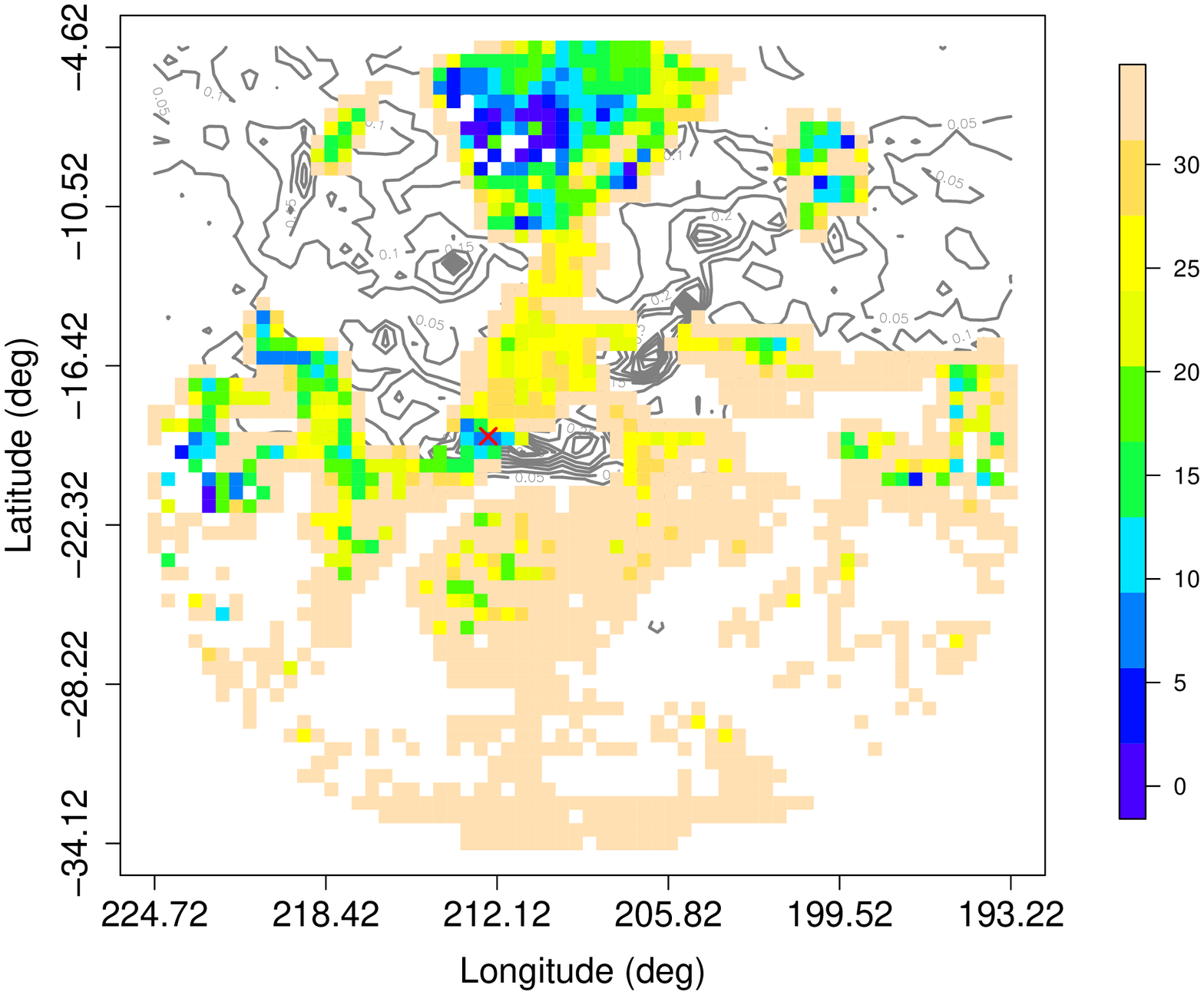}
  \includegraphics[width=\columnwidth]{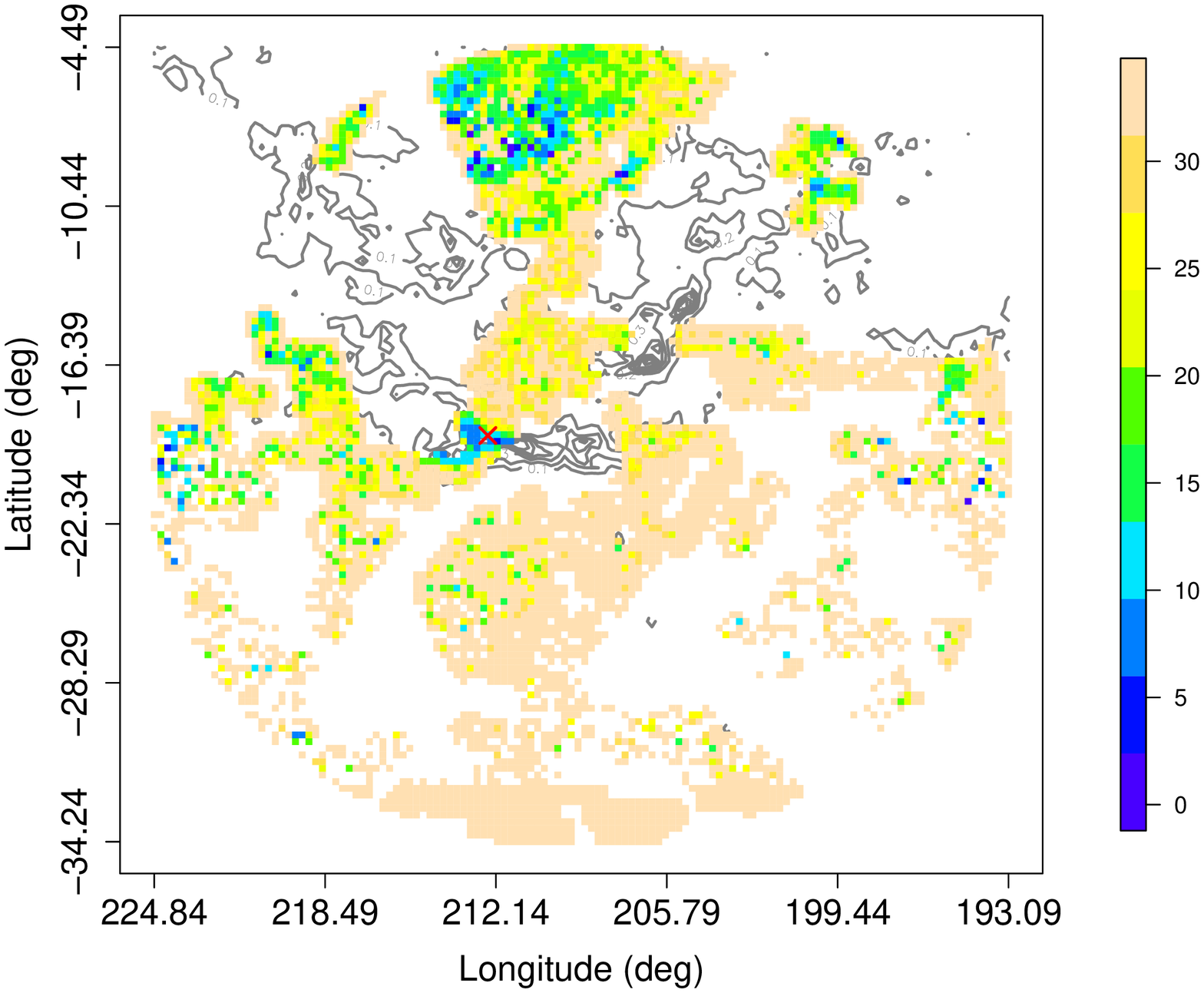}
  \includegraphics[width=\columnwidth]{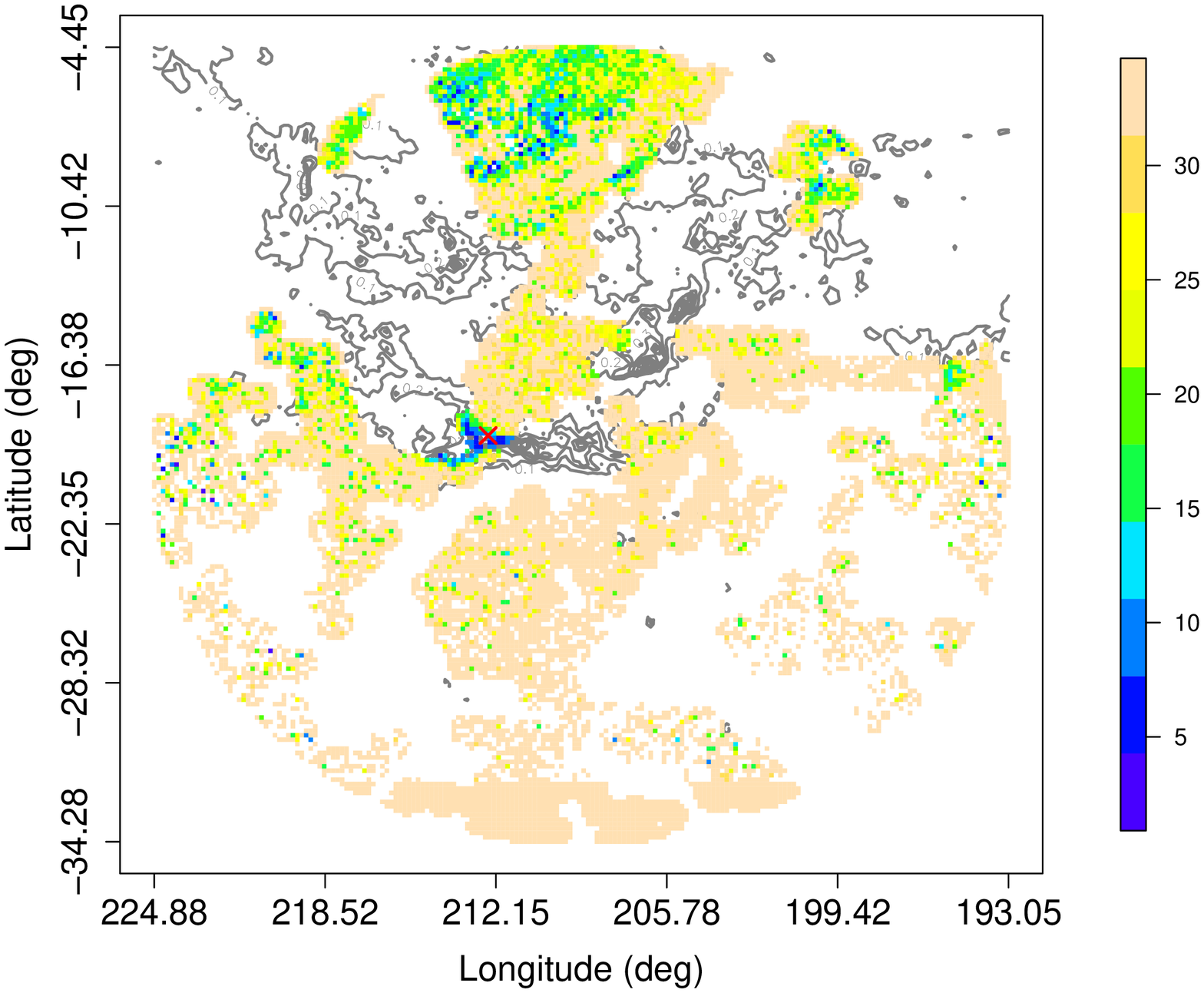}
  \caption{Orion relative extinction maps $A_{\rm NUV}/A_{\rm K}$  with
    pixel sizes of 30arcmin (up), 15arcmin (middle) and 10arcmin (down).
      Contours correspond to F07's infrared extinction map. Low
    $A_{\rm NUV}/A_{\rm K}$ values (dark color) indicate a decrease of the strength of
    the 2175$\text{\AA}$ bump and a lack of small dust particles. The
    saturation limit has been set to $A_{\rm NUV}/A_{\rm K} = 33$ (light color),
    the mean value for the ISM (GdC2015).
    The location of HD 38023, foreground to the cloud,
    is marked with a cross (see Sec. \ref{sec:rosette}).}
\label{fig:Orion_panel}
\end{figure}

\section{Results: Evidence of dust nucleation in Orion}\label{sec:results}


As Orion is located at $d\sim 400$pc, the statistics are
not good enough for 
resolving small cores and filaments in the extinction maps; the typical size 
of cores is  0.1-0.4pc that would correspond to pixel sizes of 0.9-3.4 arcmin. 
Nevertheless, we can study the large-scale distribution
of filaments and condensations.

Low values of $A_{\rm NUV}/A_{\rm K}$ point out a decrease of the strength
of the $2175\text{\AA}$ feature and a lack of small dust particles. This 
could  be produced either
by dust coagulation as observed in molecular clouds 
(\cite{Flagey09}), or by
 photo evaporation of small dust grains in shocks or heavily 
irradiated environments. 

The resolution of the 30' map  is rather rough; however, it is feasible to
identify the key areas.  Quite low $A_{\rm NUV}/A_{\rm K}$ values are found  around
$l_{gal} \sim 212$, $b_{gal} \sim -8$. This structure is partially resolved in the 15arcmin map (Fig.
\ref{fig:Orion_panel}, middle) where we
can distinguish two filaments: the first one parallel to Orion B and the other
one intersecting it perpendicularly. Inside the filament we find small
cores in the 10arcmin map (Fig. \ref{fig:Orion_panel}, down)
with $A_{\rm NUV}/A_{\rm K}$ values close to zero. Dust nucleation is most likely
to be the responsible of this $A_{\rm NUV}/A_{\rm K}$ drop.\par

Also, the area within the Orion A filament mapped  by GALEX 
presents small $A_{\rm NUV}/A_{\rm K}$ values, which is consistent with the
high density of gas, favouring the growth of dust grains. 
In Fig. \ref{fig:temp_oriA} we compare our $A_{\rm NUV}/A_{\rm K}$ map around the Orion A
molecular cloud with the optical-depth-temperature map derived by
\cite{Lombardi2014} using Herschel, Planck and IRIS data,
where color represents the dust effective
temperature (red $T\leq 12K$, blue $T\geq 30K$) and the
optical-depth is proportional
to intensity. We can see that there is an overdensity of YSO candidates
(S2014)
in the portion of the Orion A cloud that has been observed and which corresponds
to the densest region (the lowest $A_{\rm NUV}/A_{\rm K}$ values). We can also see
a filament of moderate $A_{\rm NUV}/A_{\rm K}$ values perpendicular to the
Orion B cloud, with
some slightly dense cores ($A_{\rm NUV}/A_{\rm K} \sim 20$) along it. At
$l_{gal} \sim 206\degree$, $b_{gal} ~ -19\degree$ 
we apparently find another filament with several YSO candidates placed throughout it.

\begin{figure*}
  \includegraphics[width=\textwidth]{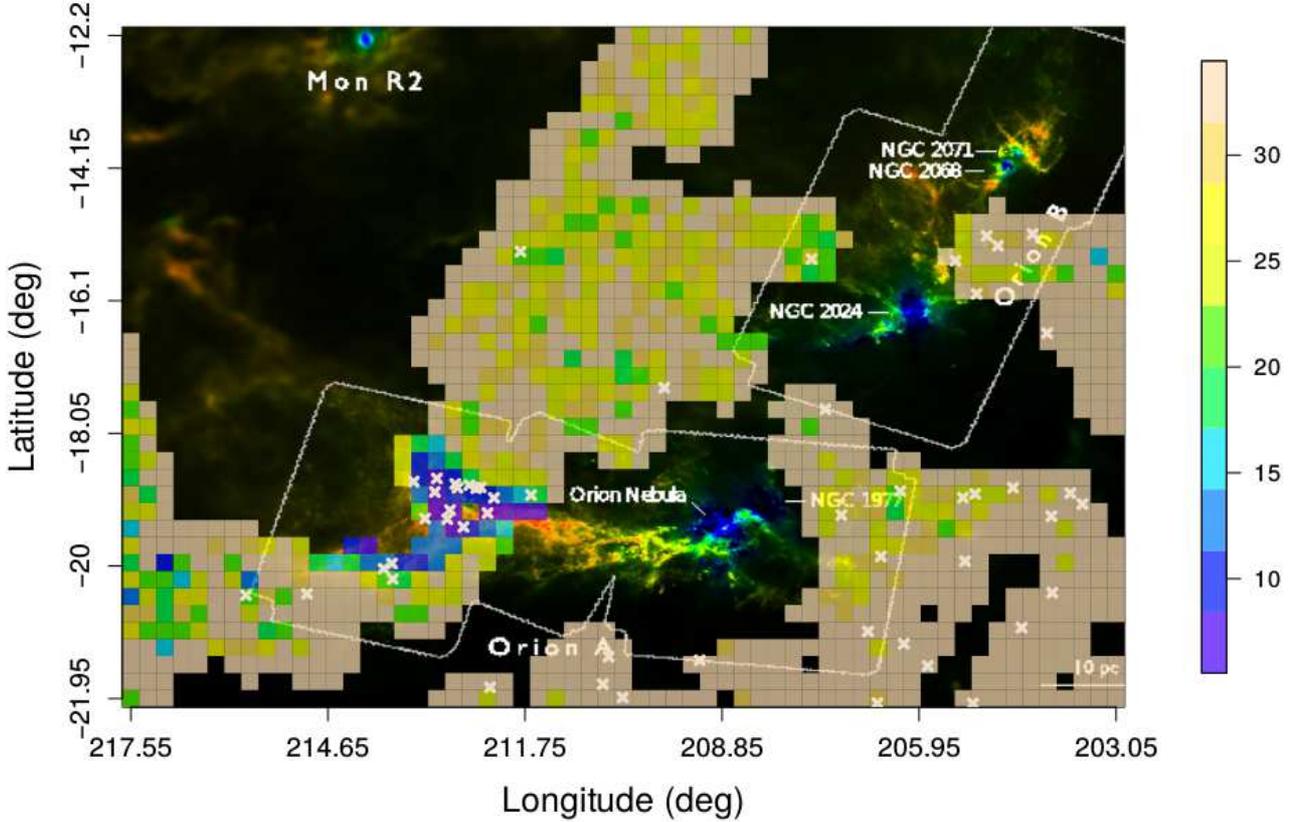}
  \caption{Orion A $A_{\rm NUV}/A_{\rm K}$ 15arcmin map
    combined with optical-depth-temperature
    map by Lombardi et al. (2014). Colors in Lombardi et al. (2014) map represent
    the dust effective temperature (red $T\leq 12K$, blue $T\geq 30K$) and the
    optical-depth is proportional
    to intensity. We can observe a decrease of the UV bump (low $A_{\rm NUV}/A_{\rm K}$ values) over the
    Orion A filament, where there exists an overdensity of YSO candidates (white crosses).}
\label{fig:temp_oriA}
\end{figure*}

At greater galactic longitudes we find intermediate $A_{\rm NUV}/A_{\rm K}$ values coinciding with
the edges of the cloud.\par

In the borders of the $\lambda$ Orionis bubble  ($l_{gal}\sim 200\degree$,
$b_{gal}\sim -8\degree$ and $l_{gal}\sim 194\degree$,
$b_{gal}\sim -16\degree$) there are
significant lower values of $A_{\rm NUV}/A_{\rm K}$ than in its
surroundings. These variations of the
  strength of the bump could be explained as the result of destruction
  of PAHs due to the strong UV radiation coming
  from the massive stellar cluster inside the bubble. The areas
mapped by GALEX correspond roughly to 
clouds L1599 and B223 (see Fig. \ref{fig:lori_combined}).

\begin{figure*}
  \includegraphics[width=\textwidth]{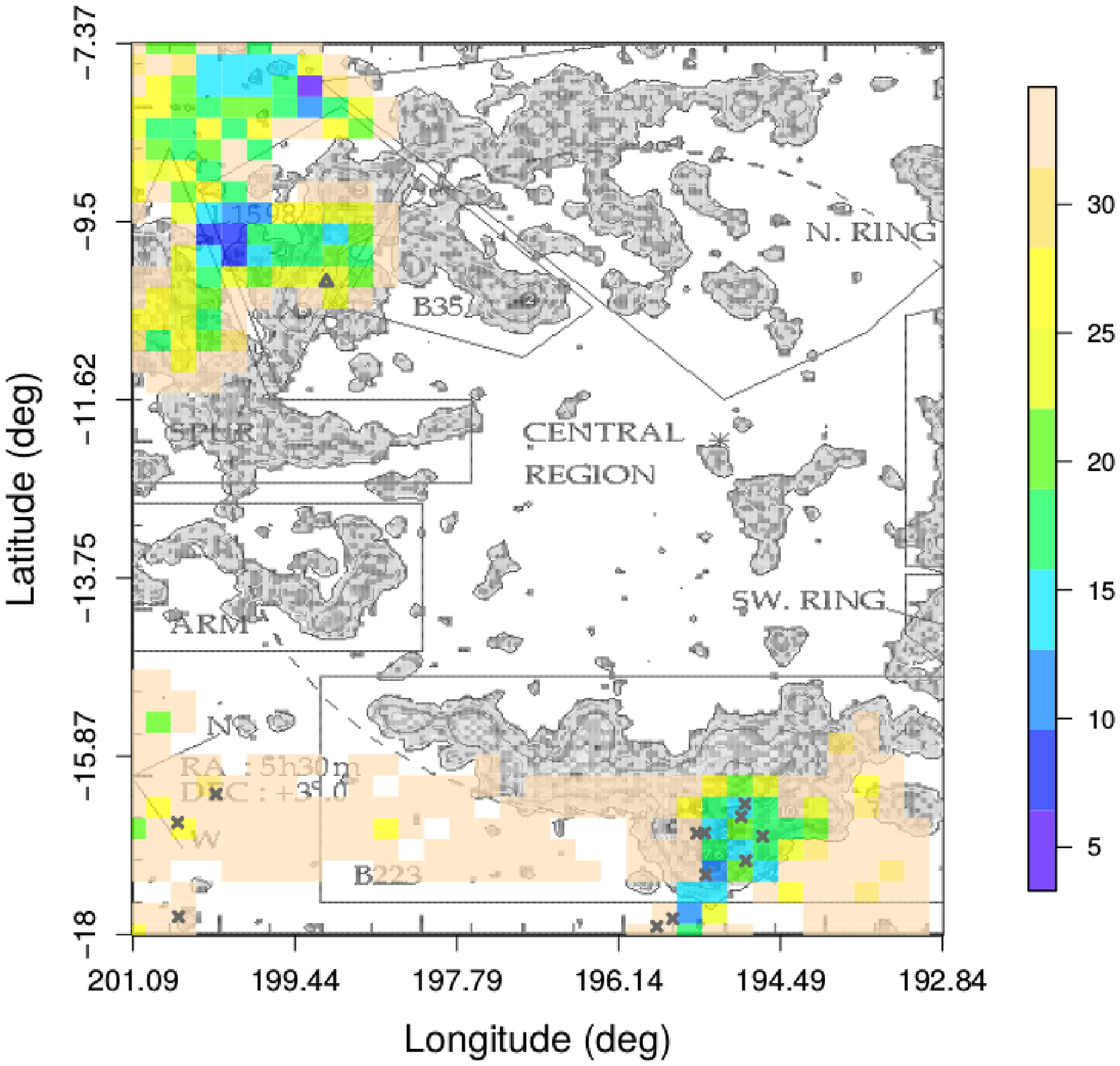}
  \caption{$\lambda$ Orionis $A_{\rm NUV}/A_{\rm K}$ 15 arcmin map
    combined with CO map by
    Lang et al. (2000).  In the area
      observed by GALEX we find
      smaller $A_{\rm NUV}/A_{\rm K}$ values than in the ISM; this fact
      could be explained as a destruction of PAHs or dust grains
    due to the UV radiation coming from the stellar cluster. Areas of special
    interest are B223 at the bottom
    of the image, where there is a gathering of YSO (crosses) and
    L1599 cloud, marked with a triangle.  A small, dense core is found
  in the outer part of the rim, near L1599 cloud.}
\label{fig:lori_combined}
\end{figure*}

Despite the fact that in L1599 our relative extinction values
are not remarkably low, they are smaller than those found in the
ISM as we are mapping the molecular material that is being
swept away inside the bubble. We find a dense
core in the outer part of the rim where
dust coagulation is  presumably taking place.

Computed $A_{\rm NUV}/A_{\rm K}$ values match almost perfectly
 with CO emission from dense, massive dark globules
 found by \cite{Lang2000} in B223; our statistics
 predict that intermediate
 size grains are forming inside these clouds. Moreover, most of
 YSO candidates selected
 by S2014 overlap with these cores, which
 could be an indicative of
triggered star formation.

\par
Finally, at $l_{gal}\sim 221\degree$, $b_{gal}\sim -20\degree$
we find a small isolated but very dense region
(Fig. \ref{fig:Orion_panel}, 15 and 10 arcmin),
apparently non-extinguished in F07's map. Nevertheless,
after a carefully comparison between GALEX, Herschel and Planck images using
ESASky we find tiny molecular cores where Herschel has detected dust
grain formation.

\section{Rosette: testing spectroscopic versus statistical methods}\label{sec:rosette}

The Rosette cloud is a well studied structure in the interstellar medium 
located at $1,330\pm48$pc from the Sun (\cite{LAL11}).  Rosette Nebula is 
close to the  Galactic Plane and it is illuminated by an open cluster 
of massive stars (NGC 2244, \cite{Martins2012}),
which is the main responsible of the cloud ionisation.
It is also considered a prototype of triggered star formation and its
properties (such as dust temperature and column density) have been
recently studied by \cite{Schneider2010} and \cite{Schneider2012}
using Herschel data.\par

We have searched the catalogue of high quality spectra of non-variable 
sources in the database of the International Ultraviolet Explorer (IUE)
(\cite{StarsCatalogue}) for UV spectra of massive O-B stars in
Orion and Rosette. In Orion we only found a star, HD 38023, foreground to the
cloud ($d=311.8$ pc, see Fig. \ref{fig:Orion_panel}) but due
to geometric dilution effects it did not allow us
to test our results. However, three O-B stars were found in the IUE
catalogue with good spectra in the
line of sight to Rosette; two foreground and another
roughly the same distance as the cloud
(HD 46106, HD 46149 and HD 46056 respectively).
The strength of the bump has been assessed from spectroscopic 
fitting following FM07, $A_{\rm bump}=\frac{\pi c_{3}}{2\gamma}$
(see Table~\ref{tab:AnuvAk_spectra}).

\begin{table}
  \caption{FM07's $c_{3}$ and $\gamma$ coefficients,
    spectroscopic  and statistical  $A_{\rm bump}$
    values for OB stars in Rosette.}
  \label{tab:AnuvAk_spectra}
\begin{tabular}{cccc}
\hline
Star & HD 46106  & HD 46149 & HD 46056 \\
\hline 
d (pc) &   809.6 & 1123.1 & 1343.9 \\
$l_{gal}$  & 206:11:55 & 206:13:12 & 206:20:09\\
$b_{gal}$  & -02:05:38 & -02:02:20 &  -02:14:50\\
\hline
$c_{3}$ & $2.93\pm 0.10$ &  $3.25\pm 0.14$ & $2.87\pm 0.12$  \\
$\gamma$  &  $0.86\pm 0.01$ &  $0.90\pm 0.02$ &  $0.88\pm 0.01$\\
Spectroscopic $A_{\rm bump}$  & $5.35 \pm 0.24$  & $5.67 \pm 0.37$ &  $5.12 \pm 0.27$\\
\hline
$\big( \frac{A_{\rm NUV}}{A_{\rm K}} \big) _{\rm map}$  & 20 & 20 &  12\\
Statistical $A_{\rm bump}$  & 4.1 $\pm$ 0.3 & 4.1 $\pm$ 0.3 & 3.3 $\pm$ 0.3\\
\hline
\end{tabular}
\end{table}

We have also estimated $A_{\rm bump}$ using statistical methods
(Eq. \ref{eq:linear_relation}) since 
Rosette was mapped during the last period of GALEX's life (DR6/7, \cite{GALEX-GR6-7}):
a grand total of 22 tiles covering $\sim 78$deg$^{2}$ and containing 52,833 point-like 
NUV sources in a search radius of $2\degree$. As for Orion, a cross-correlation test with the 2MASS catalogue
has been run to ascertain that all of them are bona-fidae astronomical sources;
only $17,384$ sources had a 2MASS counterpart.
The NUV luminosity function was found to be the same that in Orion and the
galactic latitude correction was applied as in Orion.  
F07's map has been used to compute the $A_{\rm NUV}/A_K$ maps for
pixel sizes of 10 and 15 arcmin shown in Fig.~\ref{fig:Rosette_panel}.
The image shows the well know Rosette core and  filaments. 

Both HD 46106 and HD 46149 are foreground to the cloud and have a prominent bump that
agrees with the expectations of the ISM model (FM07);  notice that the strength increases
with distance, {\it i. e.}, with dust column. In the same line of sight, but
the same distance as Rosette,  both spectroscopic (HD 46056) and statistical methods provide the lowest 
value, suggesting the destruction of small dust grains in the proximity of the cloud (either through
coagulation to larger structures or by photodestruction).

To summarise, spectroscopic values are consistently higher than those derived
from statistical averages obtained over large physical regions;
the typical size of an ISM cloud is about 2-5 pc and our 10 arcmin pixel corresponds to 
3.87 pc at Rosette's distance.

\begin{figure}
  \includegraphics[width=\columnwidth]{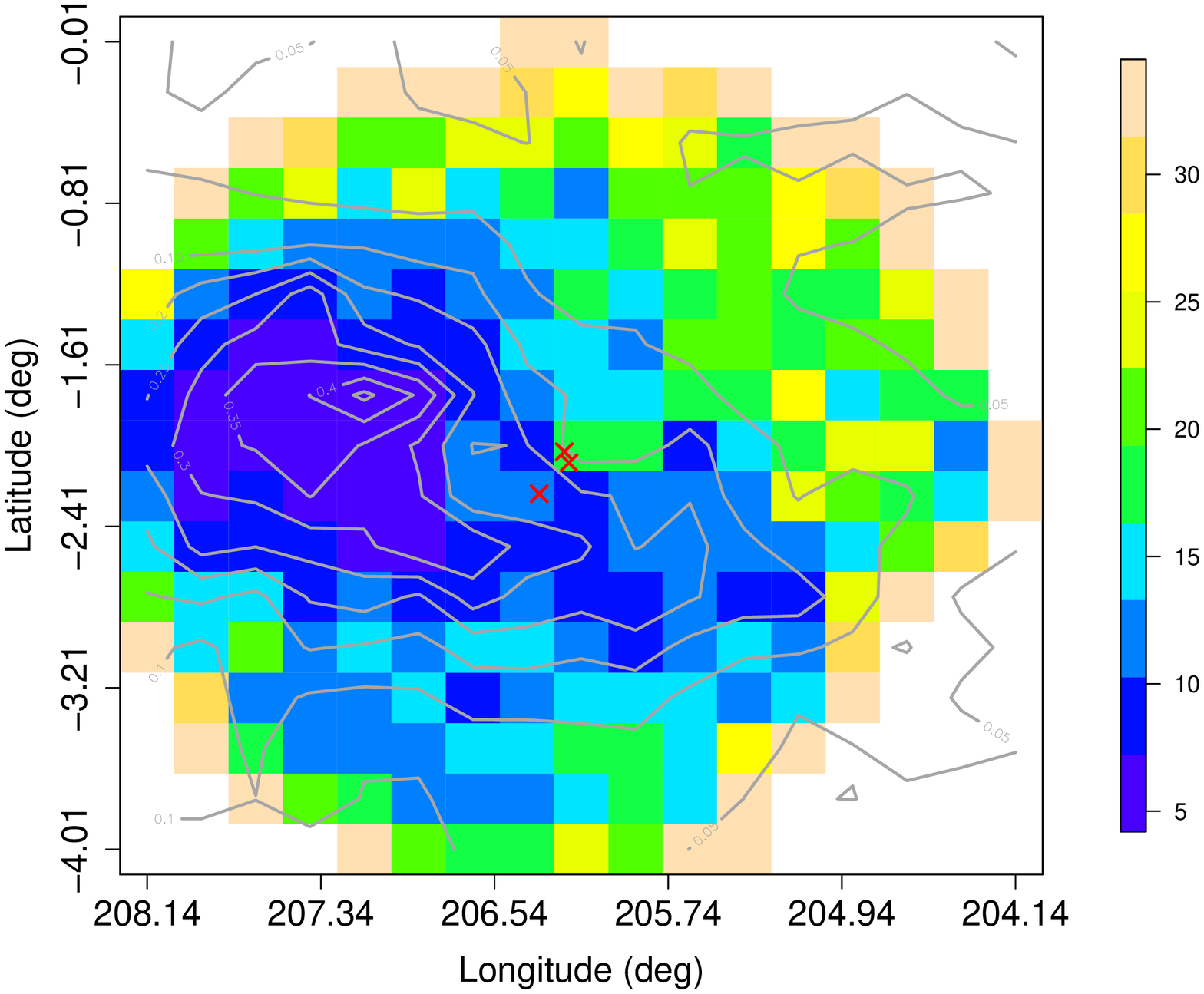}
  \includegraphics[width=\columnwidth]{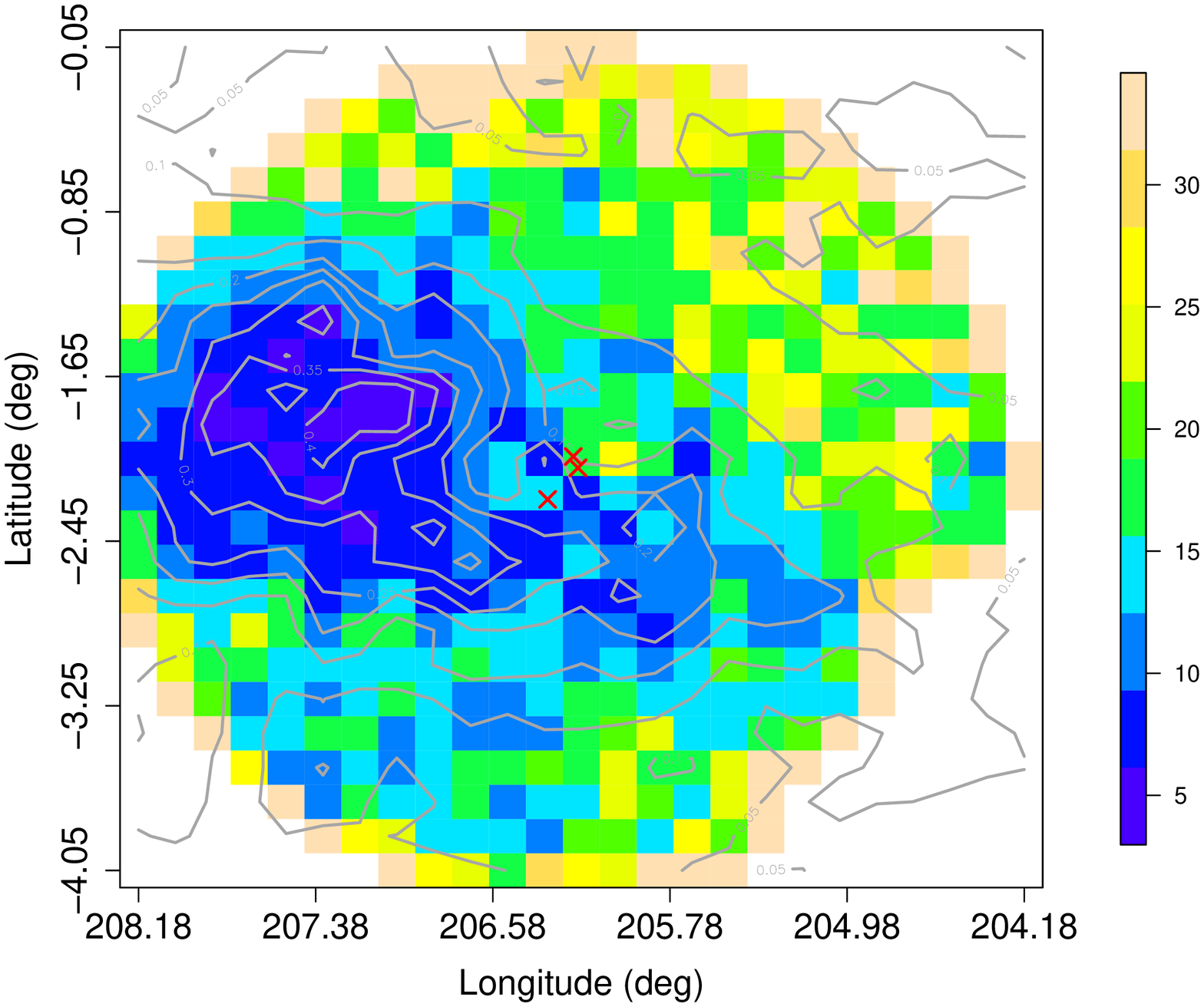}
  \caption{Rosette relative extinction map $A_{\rm NUV}/A_{\rm K}$ with
      pixel sizes of 15arcmin (up) and 10arcmin (down). Contours
      correspond to F07's infrared extinction map; 
        the colour coding is the same as in Fig \ref{fig:Orion_panel}. Low
         $A_{\rm NUV}/A_{\rm K}$ values are detected in the Rosette core.
      IUE stars HD 46023, HD 46106
      and HD 46149 are marked with a cross. Statististical $A_{bump}$ values
      for these stars, using Eq. \ref{eq:linear_relation}, are $4.1\pm 0.3$, $4.1\pm 0.1$ and $3.3\pm 0.3$
      respectively, while FM07's spectroscopic values
      ($A_{bump} = \frac{\pi c_{3}}{2\gamma}$) are $5.35\pm 0.24$,
      $5.67\pm0.37$ and $5.12\pm 0.27$.}
\label{fig:Rosette_panel}
\end{figure}

\section{Conclusions}\label{sec:conclusions}
In this work we have applied the method developed by GdC2015
to study the distribution and properties of dust in 
the ISM and molecular clouds through UV and near
infrared relative extinction maps. The main
conclusions that we derive are:

\begin{itemize}
\item There are not significant variations in the NUV
luminosity function in the Galactic Anticentre, and we
have obtained an average value consistent with
the value derived by GdC2015 in Taurus.
\item Variations of the
strength of the $2175~\text{\AA}$ bump in
relative extinction maps reveal the growth of dust
grains in some dense areas, like Orion A,  as well as
  a possible destruction of PAHs in the $\lambda$ Orionis molecular
ring.
\item Spectroscopic data provide $A_{\rm NUV}/A_{\rm K}$ values
  slightly higher than those obtained by
  statistical methods because of the
large pixel sizes.
\item The extinction values obtained by this method are
  statistically averaged, so they are fairly reliable when the
  molecular clouds are at a moderate distance like Orion but
  they are diluted at larger distances (as in Rosette).
  
\end{itemize}
\section*{Acknowledgements}
We would like to thank the referee for his/her comments that
substantially improved our manuscript.
This work has been funded by the Ministry of Economy and
Competitiviness of Spain through grants: ESP2014-54243-R and
ESP2015-68908-R. Leire
Beitia-Antero acknowledges the
receipt of a ``Beca de Excelencia'' from the
Universidad Complutense de Madrid. \\
This work has made use of data from the European Space Agency (ESA)
mission {\it Gaia} (\url{http://www.cosmos.esa.int/gaia}), processed by
the {\it Gaia} Data Processing and Analysis Consortium (DPAC,
\url{http://www.cosmos.esa.int/web/gaia/dpac/consortium}). Funding
for the DPAC has been provided by national institutions, in particular
the institutions participating in the {\it Gaia} Multilateral Agreement.



\bibliographystyle{mnras}
\bibliography{references} 




\appendix

\section{Galactic latitude correction}\label{ap:galactic_correction}

The correction factor has been evaluated for all pixel sizes,
10, 15 and 30 arcmin. In the process, evidence was found of a
real variation of the data with galactic latitude (but not with
longitude) and thus  a parametric fit could not be performed.\par
Hence, $dA_{\rm NUV}$ values were fit as a linear function of $b_{gal}$,
$dA_{\rm NUV}=m+n\times b_{gal}$,
using the Theil-Sen method (see Figure \ref{fig:dAnuv_fits}).
$dA_{\rm NUV}$, $m$ and $n$, together with residual standard error ($S_{r}$)
of the fittings are given in Table \ref{tab:dAnuv_coefs} for the three
pixel sizes considered. $N^{*,b0}_{\rm NUV}$ was measured in the 
$l_{gal}\sim 214\degree.745$, $b_{gal}\sim -8\degree$, field.

\begin{table}
\caption{Theil-Sen linear fitting results for the galactic
  latitude correction term. \textit{m} denotes the slope,
    while \textit{n} is the intercept.}
\label{tab:dAnuv_coefs}
\begin{tabular}{cccccc}
\hline
Map & $m$ & $err_{m}$ & $n$ & $err_{n}$ & $S_{r}$ \\
\hline
30arcmin & -0.075 & 0.045 & -0.307 & 0.192 & 0.218 \\
15arcmin & -0.082 & 0.056 & -0.421 & 0.223 & 0.250 \\
10arcmin & -0.073 & 0.042 & -0.401 & 0.192 & 0.220 \\
\hline
\end{tabular}
\end{table}

\begin{figure}
  \includegraphics[width=\columnwidth]{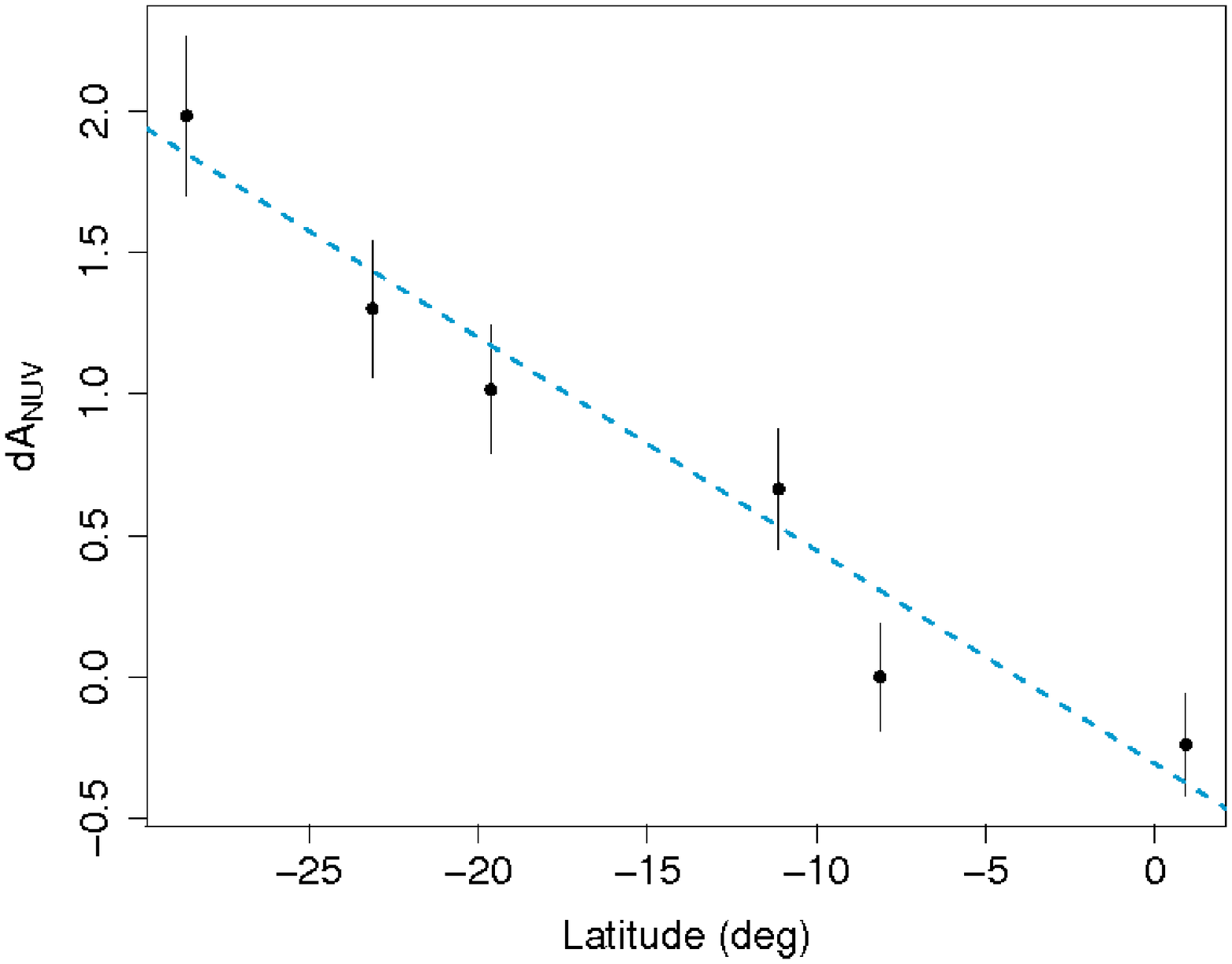}
  \includegraphics[width=\columnwidth]{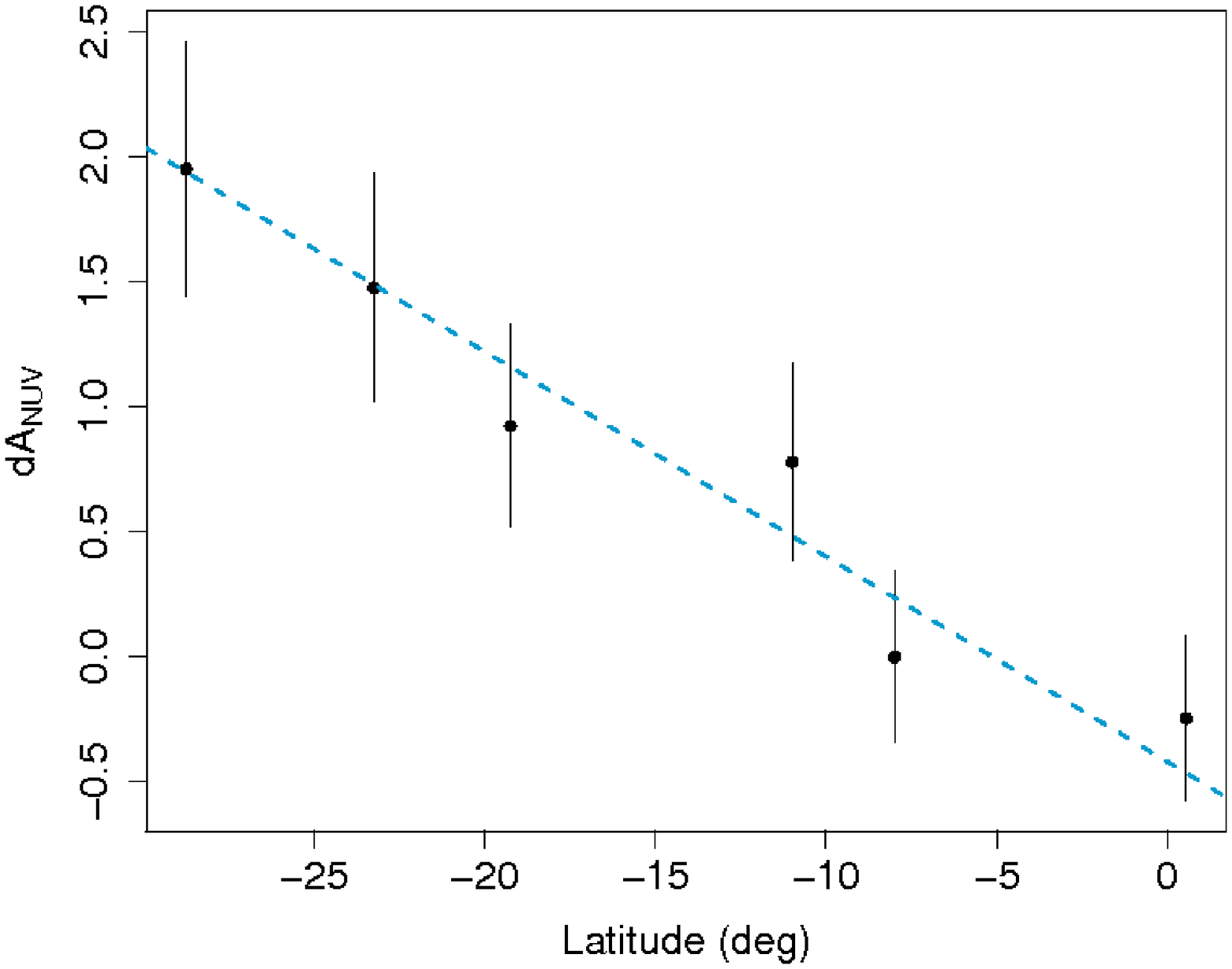}
  \includegraphics[width=\columnwidth]{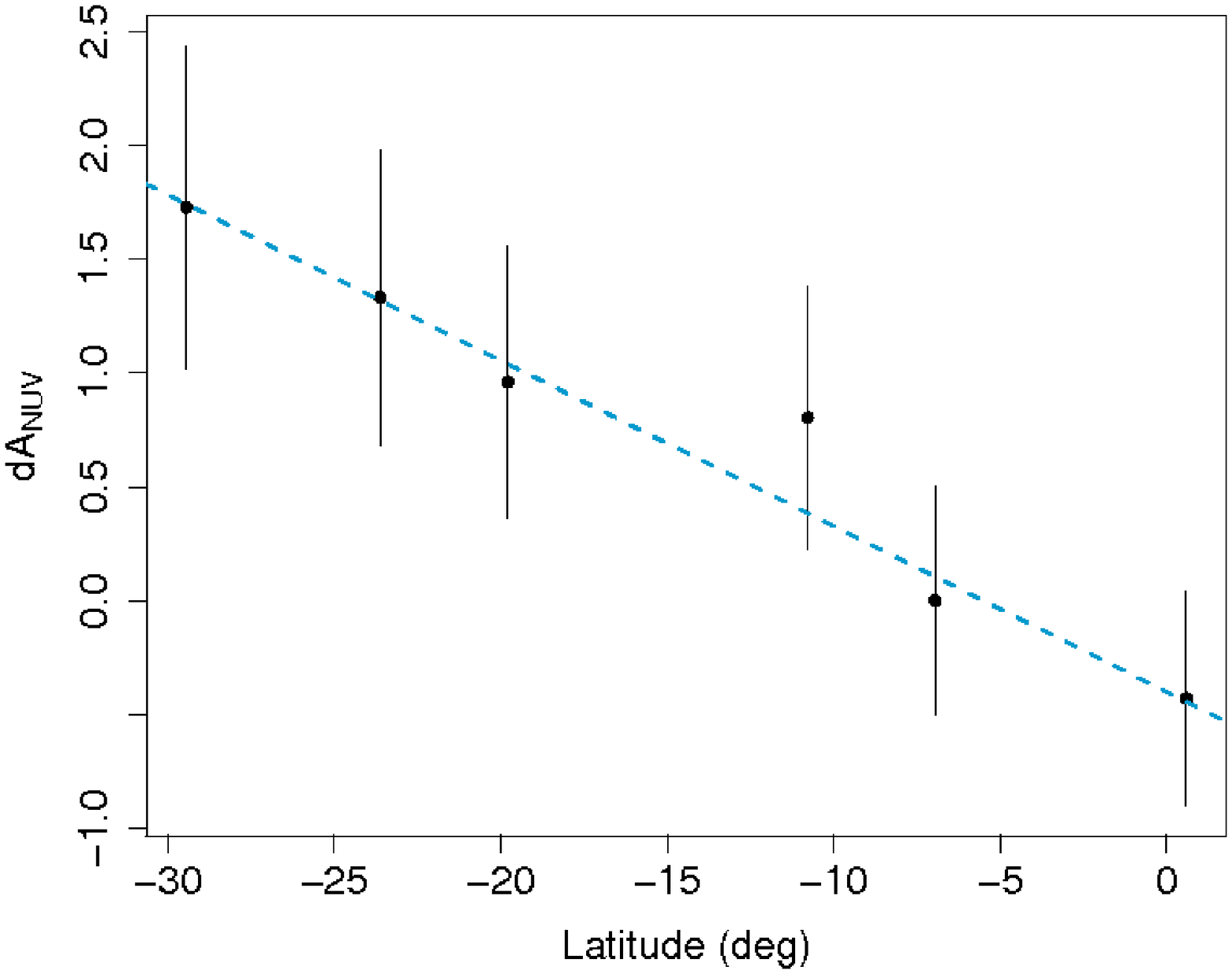}
  \caption{Theil-Sen linear fitting curves (top: 30arcmin;
    middle: 15arcmin; bottom: 10arcmin) for the galactic
    latitude correction term $dA_{\rm NUV}$. In all
      cases, the reference value was considered to be at
      $l_{gal}\sim 214\degree.745$, $b_{gal}\sim -8\degree$. The dashed
    line corresponds to
    the linear fitting.}
  \label{fig:dAnuv_fits}
\end{figure}


\bsp	
\label{lastpage}
\end{document}